\documentclass[10pt,conference]{IEEEtran}

\usepackage{xcolor} 
\usepackage{ifthen} 
\usepackage{fontawesome} 

\ifthenelse{\isundefined{\editmode}}{
  \newcommand{\editnote}[3]{}
  
}{
  \newcommand{\editnote}[3]{%
    \xspace\colorbox{#1}{\sffamily \smaller \textcolor{white}{~\faCommenting{}~#2~}}%
    \textcolor{#1}{~#3}\xspace}
  
}

\definecolor{mygreen}{HTML}{02818a}
\definecolor{nord10}{HTML}{5E81AC}
\definecolor{darkorange}{HTML}{FE6741}

\newcommand{\aj}[1]{\editnote{nord10}{AJung}{#1}}
\newcommand{\lili}[1]{\editnote{cyan}{Lili}{#1}}
\newcommand{\rf}[1]{\editnote{darkorange}{Rufeng}{#1}}

\newcommand{\etal}{et al.}

\usepackage[utf8]{inputenc}
\usepackage{xcolor,colortbl}
\usepackage{graphicx}
\usepackage{color}
\usepackage{pifont}
\usepackage{ifthen}
\usepackage{multirow}
\usepackage{booktabs}
\usepackage[inline]{enumitem}
\usepackage[ruled,linesnumbered]{algorithm2e}
\usepackage{tcolorbox}
\usepackage{booktabs}
\usepackage{tabularx}
\newcolumntype{L}{>{\raggedright\arraybackslash}X}%
\newcolumntype{R}{>{\raggedleft\arraybackslash}X}%
\newcolumntype{C}{>{\centering\arraybackslash}X}%
\usepackage{xspace}
\usepackage{fontawesome}
\usepackage{balance}
\usepackage{listings}
\usepackage{subcaption}
\usepackage{url}
\usepackage{tikz}
\usetikzlibrary{trees}
\usepackage{forest}
\usepackage{float}
\usetikzlibrary{arrows.meta}
\usepackage{amssymb}
\lstdefinelanguage{JavaScript}{
  keywords={const, let, async, await, typeof, new, true, false, catch, function, return, null, try, catch, switch, var, if, in, while, do, else, case, break},
  ndkeywords={class, export, boolean, throw, implements, import, this},
  sensitive=false,
  comment=[l]{//},
  morecomment=[s]{/*}{*/},
  morestring=[b]',
  morestring=[b]"
}

\usepackage{titlesec}
\titleformat*{\paragraph}{\bfseries}
\IEEEoverridecommandlockouts

\begin{document}

\title{Examining the Usage of Generative AI Models in Student Learning Activities for Software Programming}

\makeatletter
\newcommand{\linebreakand}{%
  \end{@IEEEauthorhalign}%
  \hfill\mbox{}\par
  \mbox{}\hfill\begin{@IEEEauthorhalign}%
}
\makeatother

\author{%
\IEEEauthorblockN{Rufeng Chen\textsuperscript{\dag}}
\IEEEauthorblockA{\textit{McGill University}\\
Montreal, Canada\\
rufeng.chen@mail.mcgill.ca}
\and
\IEEEauthorblockN{Shuaishuai Jiang\textsuperscript{\dag}}
\IEEEauthorblockA{\textit{McGill University}\\
Montreal, Canada\\
shuaishuai.jiang@mail.mcgill.ca}
\and
\IEEEauthorblockN{Jiyun Shen}
\IEEEauthorblockA{\textit{Carnegie Mellon University}\\
Pittsburgh, USA\\
jiyuns@andrew.cmu.edu}
\linebreakand
\IEEEauthorblockN{AJung Moon}
\IEEEauthorblockA{\textit{McGill University}\\
Montreal, Canada\\
ajung.moon@mcgill.ca}
\and
\IEEEauthorblockN{Lili Wei}
\IEEEauthorblockA{\textit{McGill University}\\
Montreal, Canada\\
lili.wei@mcgill.ca}
\thanks{\textsuperscript{\dag}These authors contributed equally to this work.}
}

\maketitle

\begin{abstract}
    The rise of Generative AI (GenAI) tools like ChatGPT has created new opportunities and challenges for computing education. 
Existing research has primarily focused on GenAI's ability to complete educational tasks and its impact on student performance, often overlooking its effects on knowledge gains. 
In this study, we investigate how GenAI assistance compares to conventional online resources in supporting knowledge gains across different proficiency levels. 
We conducted a controlled user experiment with 24 undergraduate students of two different levels of programming experience (beginner, intermediate) to examine how students interact with ChatGPT while solving programming tasks. 
We analyzed task performance, conceptual understanding, and interaction behaviors. 
Our findings reveal that generating complete solutions with GenAI significantly improves task performance, especially for beginners, but does not consistently result in knowledge gains.
Importantly, usage strategies differ by experience: beginners tend to rely heavily on GenAI toward task completion often without knowledge gain in the process, while intermediates adopt more selective approaches. 
We find that both over-reliance and minimal use result in weaker knowledge gains overall. 
Based on our results, we call on students and educators to adopt GenAI as a learning rather than a problem solving tool. 
Our study highlights the urgent need for guidance when integrating GenAI into programming education to foster deeper understanding.
\end{abstract}

\section{Introduction}
\label{sec:introduction}
The rapid development of Generative Artificial Intelligence (GenAI) has led to its widespread adoption across various domains to boost productivity and streamline workflows. 
Large Language Models (LLMs), such as OpenAI's ChatGPT and Codex, Google Gemini, and GitHub Copilot, have been integrated into domains including software engineering ~\cite{2021Chen,10.1145/3491101.3519665}, healthcare ~\cite{Singhal2023}, education ~\cite{KASNECI2023102274}, creative writing ~\cite{10.1145/3544548.3581225,reif2022recipearbitrarytextstyle}, and digital music ~\cite{liu2023generativediscotexttovideogeneration}, offering capabilities such as code generation, question answering, and image generation.

Existing research explored various aspects of GenAI integration in education. 
Some studies evaluated GenAI’s performance on programming tasks~\cite{10.1145/3593663.3593695}, user interface design education~\cite{10.1145/3615335.3623035}, and computer vision coursework~\cite{10.1145/3636243.3636263}. 
Others focused on assessing the accuracy and usability of GenAI-generated responses~\cite{10.1145/3511861.3511863, 10.1145/3576123.3576134}. 
Additionally, researchers proposed GenAI-powered tools to support student learning and instruction~\cite{10.1145/3626252.3630938, 10.1145/3636243.3636252}.
Collectively, these studies highlight GenAI’s strong potential to assist in completing educational tasks and improving student performance.

Despite this progress, many educators remain cautious about incorporating GenAI tools into computing curricula. 
A key source of concern is the lack of clarity on how to assess students' knowledge gains when GenAI is involved. 
These concerns span from potential academic dishonesty ~\cite{LEE2024100253} to uncertainty about whether students genuinely understand the material or are simply completing tasks with AI assistance. 
While some educators have begun experimenting with GenAI in classrooms, much of the current work ~\cite{bhalerao, 10.1145/3638067.3638100, 10.1145/3622780.3623648, 10.1145/3613904.3642706} has focused on prompt engineering, evaluating GenAI-generated content, and measuring task performance, rather than on whether students achieve meaningful knowledge gains.

Moreover, limited research examined how students with varying proficiency levels engage with GenAI during learning activities. 
Existing research often focuses on students from a single course~\cite{10.1145/3643795.3648379} or cohort~\cite{11028221}, overlooking variations in proficiency levels.
It remains unclear whether students with different programming backgrounds use GenAI differently, and how such usage patterns influence both their task performance and knowledge gains.

To address this gap, our study investigates how GenAI assistance compares to conventional online resources in supporting knowledge gains across different proficiency levels. 
Specifically, we pose the following research questions:
\begin{itemize}[noitemsep, topsep=1pt, leftmargin=*]
    \item \textbf{RQ1:} How do students perceive the use of GenAI for assisting with programming tasks?
    \item \textbf{RQ2:} How does access to GenAI impact task performance for students with different proficiency levels?
    \item \textbf{RQ3:} How does access to GenAI impact knowledge gains for students with different proficiency levels?
    \item \textbf{RQ4:} How do students interact with GenAI when solving programming tasks, and how do these interaction patterns vary by proficiency level?
\end{itemize}
\aj{Add an explicit declaration of our research question here -- then we can frame the hypotheses in the experiment methods/analysis sections accordingly.}\rf{revised}
We conducted a controlled user study involving 24 undergraduate students, comparing how beginner and intermediate learners interacted with ChatGPT to solve a programming task. 
Participants were divided into four groups based on their proficiency level and whether they had access to GenAI. 
We analyzed their task performance, knowledge gains, and interaction behaviors to explore how different usage strategies shaped their learning experiences.

Our study reveals several key findings:
\begin{itemize}[noitemsep, topsep=1pt, leftmargin=*]
    \item Using GenAI for complete solution generation boosted task performance, but did not lead to improvements in knowledge gains, especially for students who focused solely on task completion.
    \item While GenAI helped beginners overcome implementation barriers, it did not enhance their conceptual understanding. 
    In contrast, intermediates who complete the task without GenAI demonstrated stronger comprehension of both solutions and underlying concepts.
    \item The timing of a student's exposure to GenAI relative to their acquisition of programming skills matters. \lili{It is a bit confusing to see ``how students FIRST ENCOUNTERED it''. What does it mean?}\rf{revised}
    Beginners, who were more likely to be exposed to GenAI prior to developing their own programming skills, tended to treat it as a primary problem-solving tool and had limited experience with conventional online resources. 
    In contrast, intermediates, who learned programming before engaging with GenAI, used it more cautiously.
    \item Both over-reliance and underuse of GenAI led to weaker knowledge gains, highlighting that most participants lacked effective learning strategies with GenAI.
\end{itemize}

Building on our results, we provide practical recommendations for both students and educators on how GenAI can be effectively integrated into classrooms to support learning and task completion.

All study materials, datasets, and analysis scripts are available on GitHub~\cite{replication_package}.

\section{Related Work}
\label{sec:related-work}

The growing integration of GenAI has sparked significant interest in educational applications, especially in computer science education. 
Existing research has largely focused on evaluating the capabilities of these models to solve programming problems and analyzing how students interact with GenAI tools during learning activities.

\textbf{GenAI's Problem-Solving Capabilities.}
Several studies have examined the ability of GenAI models to solve a range of programming tasks. 
These include introductory programming exercises~\cite{10.1145/3511861.3511863}, object-oriented programming assignments~\cite{10.1145/3576123.3576134}, data structures and algorithms exams~\cite{10.1145/3587102.3588814}, and other programming tasks~\cite{10.1145/3545945.3569770, 10.1145/3587102.3588805}.
For example, Dakhel \etal{}~\cite{dakhel2023githubcopilotaipair} compared the quality of Codex-generated code to student-written code. 
These findings support the view that GenAI tools are capable of effectively solving complex programming problems, often rivaling student performance.

While most prior work has emphasized correctness and code quality as evaluation metrics, our work focuses on understanding whether GenAI can also support students in learning programming concepts, not just solving problems. 
This distinction is essential for assessing the educational value of GenAI in a learning context.

\textbf{Student–GenAI Interaction and Learning Behaviors.}
Other studies have explored how students interact with GenAI tools during programming tasks. 
For example, Promptly \etal{}~\cite{denny2023promptlyusingpromptproblems} analyzed how 54 students wrote prompts to solve CS1-level problems with GenAI. 
Nguyen \etal{}~\cite{10.1145/3613904.3642706} studied how 120 beginning coders approach writing and editing prompts.
Rahe and Maalej~\cite{10.1145/3715762} studied the interaction patterns of 37 programming students using ChatGPT to complete a coding exercise. 
Similarly, Prather \etal{}~\cite{10.1145/3617367} examined how 19 students used GitHub Copilot to develop a Minesweeper game as a final project in an introductory CS course. 
These studies found that many students had difficulty writing helpful prompts, which limited their ability to solve programming tasks using GenAI.
\aj{Have a sentence here that summarizes what these three articles found about GenAI use in student learning.}\rf{revised}

Kazemitabaar \etal{}~\cite{10.1145/3544548.3580919} conducted a comparative study involving 33 students with Codex and 36 who completed the same 45 programming tasks without GenAI support.
Likewise, Sivasakthi and Meenakshi~\cite{Sivasakthi2025} evaluated the task performance of 160 students enrolled in a Python course, comparing a ChatGPT-assisted group with a control group.
Both studies found that using GenAI tools can positively impact performance on programming tasks.
\aj{Same thing here. Describe what they found from these studies that is relevant to this paper.}\rf{revised}

However, most of these works focused on how students used GenAI and what outputs were generated, without thoroughly investigating how the GenAI use affects students' learning. 
In contrast, our work examines not only task performance, but also student knowledge gains, specifically, whether interacting with GenAI helps them understand programming concepts more effectively and deeply.

Penney \etal{}~\cite{10.1145/3719160.3736625} provide a comparison by evaluating the effectiveness of GenAI versus human tutors in guiding 20 students enrolled in an introductory programming course. 
Similarly, Liu \etal{}\cite{10430054} evaluated the use of ChatGPT in comparison to Stack Overflow for helping students with programming tasks.
These studies analyzed the advantages and disadvantages of GenAI for beginners or single-cohort students, but they overlooked how differences in proficiency level influence its impact.
\aj{Add a sentence here on what Penney and Liu work found in their results and transition to the next sentence.}\rf{revised}
In our study, we extend this comparison by evaluating students who used GenAI and those who relied solely on conventional online resources.
These resources include Stack Overflow, official documentation, and other non-AI websites, offering a realistic representation of common learning environments without GenAI support.
Furthermore, we analyze how students with different levels of programming proficiency, beginner and intermediate, differ in their usage patterns and knowledge gains when assisted by GenAI.
\section{Experiment Setup}
\label{sec:experiment}

To evaluate the impact GenAI has on student learning of programming skills, we conducted a $2 \times 2$ between-subjects experiment with 24 participants: (\textit{beginner vs. intermediate programming experience}) $\times$ (\textit{with vs. without ChatGPT}).  
We sought to simulate a self-directed learning scenario in programming education, where a student learns a concept and then applies it to solve a relevant programming task using a single external aid—in this case, ChatGPT.

We recruited undergraduate students from a Canadian university via emails and posters in an introductory and an intermediate-level programming courses. 
Participation was voluntary.
All participants received \$30 CAD for their time regardless of their performance on the experimental task.

Participants were classified as intermediate if they met either of the following criteria: (1) enrolled in a third or fourth-year software engineering or computer science program or (2) reported 3+ years of coding experience.
This distinction reflects their typical academic exposure, with Beginners having completed only introductory programming and data structure courses, and Intermediates having undertaken advanced courses such as algorithms and software engineering.\rf{revised}
Twelve beginner and twelve intermediate programmer participants were then equally divided between the two experimental conditions -- complete programming tasks with or without ChatGPT -- resulting in six participants per randomized block.
\aj{This would be a great place to list the hypotheses here.} This design allows us to compare the impact of ChatGPT access not only on overall task performance but also on how its benefits may differ between novice and more experienced learners.
In addition to performance metrics, we collected survey responses and analyzed participants’ chat histories to better understand their attitudes toward ChatGPT and the strategies they used when engaging with the tool.
The survey included multiple-choice questions on participant background, Likert-scale items assessing perceptions of ChatGPT (e.g., overall usefulness, trust, confidence),  and open-ended questions examining their understanding of the solution and related concepts.
These data enabled a deeper analysis of how students interact with generative AI during learning tasks and how such tools are perceived across different skill levels.
The study received approval of the research ethics board at the university (REB Certificate \# \texttt{23-10-018}).



\subsection{Study Procedure}

The experiment was conducted in a controlled setting and following the procedure outlined below:

\begin{itemize}[noitemsep, topsep=1pt, leftmargin=*]
    \item \textbf{Participant Onboarding and Consent:}
    All participants were first required to read and sign a consent form outlining the study’s purpose, procedures, data privacy measures, and their rights as research participants. Participation was entirely voluntary, and students could withdraw at any point without penalty.
    In this section, we emphasized that our goal was not to evaluate their programming skills, but rather to investigate how well they could learn and apply a new programming concept. 
    This clarification was intended to reduce performance pressure and encourage a focus on conceptual understanding.
    \item \textbf{Pre-Study Survey:}
    Participants completed an online pre-study survey designed to collect background information, including programming experience, prior exposure to ChatGPT or similar tools, and self-assessed proficiency.
    \item \textbf{Concept Introduction via Video:}
    All participants watched an educational video (10 minutes) introducing the Dynamic Sliding Window technique to ensure a baseline conceptual understanding before attempting the task.
    This concept is not covered in the university’s standard curriculum.
    \lili{Add a sentence to explain that this concept is not included in the curricula of the university}\rf{revised}
    \item \textbf{Coding Task:}
    After the video, participants were asked to solve a LeetCode-style programming problem that required applying the concept introduced in the video.
    Participants in the ChatGPT condition were instructed to use only GPT-3.5-turbo as their external resource during the task.
    Participants in the non-ChatGPT condition were allowed to use conventional online resources such as official documentation and websites (e.g., Stack Overflow), but not AI-based assistants.
    All participants had 30 minutes to complete the task.
    During the session, screen activity was recorded to capture interaction behavior and task progression. Test cases were provided to help students validate and debug their solutions.
    The choice of concept, task difficulty, and time limit were determined through multiple rounds of pilot study to ensure appropriateness and balance across conditions.
    In this step, we collected their code, chat history, and screen recordings to assess task performance (RQ2) and interaction patterns (RQ4).
    The Dynamic Sliding Window task was selected after several pilot studies to balance difficulty for both Beginner and Intermediate participants and to fit within the 30-minute time limit. 
    We tested multiple candidate problems (e.g., array rotation, Collatz simulation, fixed-length sliding windows) with nine students across different academic levels and found this task provided an appropriate challenge without excessive complexity.
    To avoid prior knowledge effects, we verified that none of the participants had previously encountered this concept. 
    Although using a single task may introduce validity threats, this choice ensured comparability between conditions and minimized other factors such as differences between tasks and participant fatigue.\rf{revised.}
    \item \textbf{Post-Study Survey:}
    Upon completing the coding task, participants filled out a post-study survey assessing:
    Their understanding of the concept.
    Explanation of their code.
    Satisfaction with ChatGPT(if available).
    Attitudes towards the use of generative AI for learning.
    In this step, we used multiple-choice questions to gather participants’ perceptions of ChatGPT (RQ1), and open-ended questions to assess their knowledge gains (RQ3).
\end{itemize}

\subsection{Measures}
\aj{Either here or after the Study Procedure, I would add a subsection on "Measures" here and articulate all the items we measure and report as results. Indicate the type of questions in the questionnaire (refer to Fig 2 for full list of questions to reduce redundancy). The paragraph headings of 4.2 would be listed her and refe to the question numbers -- that way, we don't need to articulate them in the Results section. Were these 5-point Likert scale measures or 7-point scales from strongly agree to strongly disagree? etc. What was measured to quantify task performance? What did we use to measure knowledge gains? etc. Right now they are mixed in the results section, making the results quite bulky.}
To evaluate the effects of ChatGPT use on task performance and knowledge gains, we collected both behavioral and self-reported data. 
This section outlines the key measures and analysis to answer the research questions.

\textbf{RQ1: Perceptions of ChatGPT.}
We included ten Likert-scale questions in the post-study survey to assess participants' perceptions of ChatGPT ( Figure~\ref{fig:perceptions}).
All questions were rated on a 5-point Likert scale ranging from \textit{Strongly Disagree} to \textit{Strongly Agree}. 
To better understand participants' experiences using ChatGPT for programming tasks, we grouped related Likert-scale items into thematic categories: Overall Acceptance and Future Use (Q1, Q2), Communication and Interaction Quality (Q3, Q4, Q9), Guidance (Q5), Syntax Support (Q6, Q8), Confidence and Trust (Q7, Q10).

\textbf{RQ2: Task Performance.}
We used Randoop~\cite{randoop} to randomly generate 500 test cases that evaluated the correctness of participants’ implementations.
To ensure the reliability of the generated cases, we first created them from a fully correct reference solution that passed all tests and then manually verified that imperfect implementations could not pass every case.
This process gave us confidence that the final test suite effectively distinguished correct from incorrect solutions.\rf{revised.}
To understand how ChatGPT impacts participants’ task performance, we categorized all 24 participants into four distinct groups based on their proficiency levels and whether they had access to ChatGPT. 
The groups are: \textit{Beginner with ChatGPT}, \textit{Beginner without ChatGPT}, \textit{Intermediate with ChatGPT}, and \textit{Intermediate without ChatGPT}. 
Each group consisted of six participants.

\textbf{RQ3: Knowledge Gains.}
To assess whether participants truly understood the programming concept and could apply it effectively, we included two reflective questions in the post-study survey:
\begin{itemize}[noitemsep, topsep=1pt, leftmargin=*]
    \item \textit{Please briefly explain your approach to solving the coding question. 
    Your answer may include a brief description of your thought process and your solution.}
    \item \textit{Please explain briefly how your solution is related to the concept covered in the video.}
\end{itemize}

The first question assesses participants’ ability to explain their reasoning and applied strategies, reflecting their understanding of the solution. 
The second question evaluates whether they could connect their implementation to the concept presented in the instructional video.
We used a close coding scheme to categorize knowledge gains into three levels: \textit{Strong Understanding}, \textit{Partial Understanding}, and \textit{Weak Understanding}.

\textbf{Solution Understanding}
\begin{itemize}[noitemsep, topsep=1pt, leftmargin=*]
    \item \textbf{Strong Understanding}: Clearly and accurately explains their code and problem-solving logic.
    For example, “I used a sliding window with two pointers… tracked min/max elements… expanded right if within the limit, otherwise moved left and reset min/max.”
    \item \textbf{Partial Understanding}: Demonstrates some understanding but includes gaps or incorrect reasoning.
    For example, "using a for loop to check the absolute difference of all elements” which shows awareness of the goal but makes the algorithm unnecessarily complex.
    \item \textbf{Weak Understanding}: Cannot explain their approach or provides an incorrect/confused explanation (e.g., “Let ChatGPT write the solution”).\rf{revised.}
\end{itemize}

\textbf{Concept Understanding}
\begin{itemize}[noitemsep, topsep=1pt, leftmargin=*]
    \item \textbf{Strong Understanding}: Accurately connects the solution to the underlying concept with depth and clarity.
    For example, “...This means that I do not explicitly cover all possible combinations to see if a valid solution exists...” which shows an understanding of the difference between a brute-force approach and the optimized dynamic sliding window solution.
    \item \textbf{Partial Understanding}: Makes a weak or vague connection to the concept.
    For example, “...I would compare each element in the window to the newest element...” which reflects awareness of the window structure but not how to effectively utilize the information stored within it.
    \item \textbf{Weak Understanding}: Fails to relate the solution to the concept or does so incorrectly (e.g., “I accidentally took a brute-force approach... I know I need to shift my window at the front and back, but I got stuck,” which shows limited awareness of how to maintain the sliding window structure).\rf{revised.}
\end{itemize}
Following these criteria, two authors independently coded all responses and then discussed any discrepancies. 
A third author was involved to resolve remaining disagreements.
We obtained Cohen's kappas of 0.81 for the first question and 0.84 for the second, indicating a high level of agreement.

\textbf{RQ4: Interaction Patterns with ChatGPT.}
To understand how participants interacted with ChatGPT, we analyzed their chat histories using an open coding approach. 
This qualitative analysis allowed us to identify recurring patterns in how students interacted with the tool.
Two researchers independently reviewed and coded the chat transcripts to identify interaction behaviors such as prompt content and code inputs.
Disagreements were resolved through discussion, and a third author was involved when consensus could not be reached. 
Through this process, we developed a coding scheme that captured common usage patterns as shown in Table~\ref{tab:chat_history}.
\section{Results}
\label{sec:results}

\subsection{Participant Demographics}

\begin{figure*}[t]
    \centering
    \includegraphics[width=0.8\linewidth]{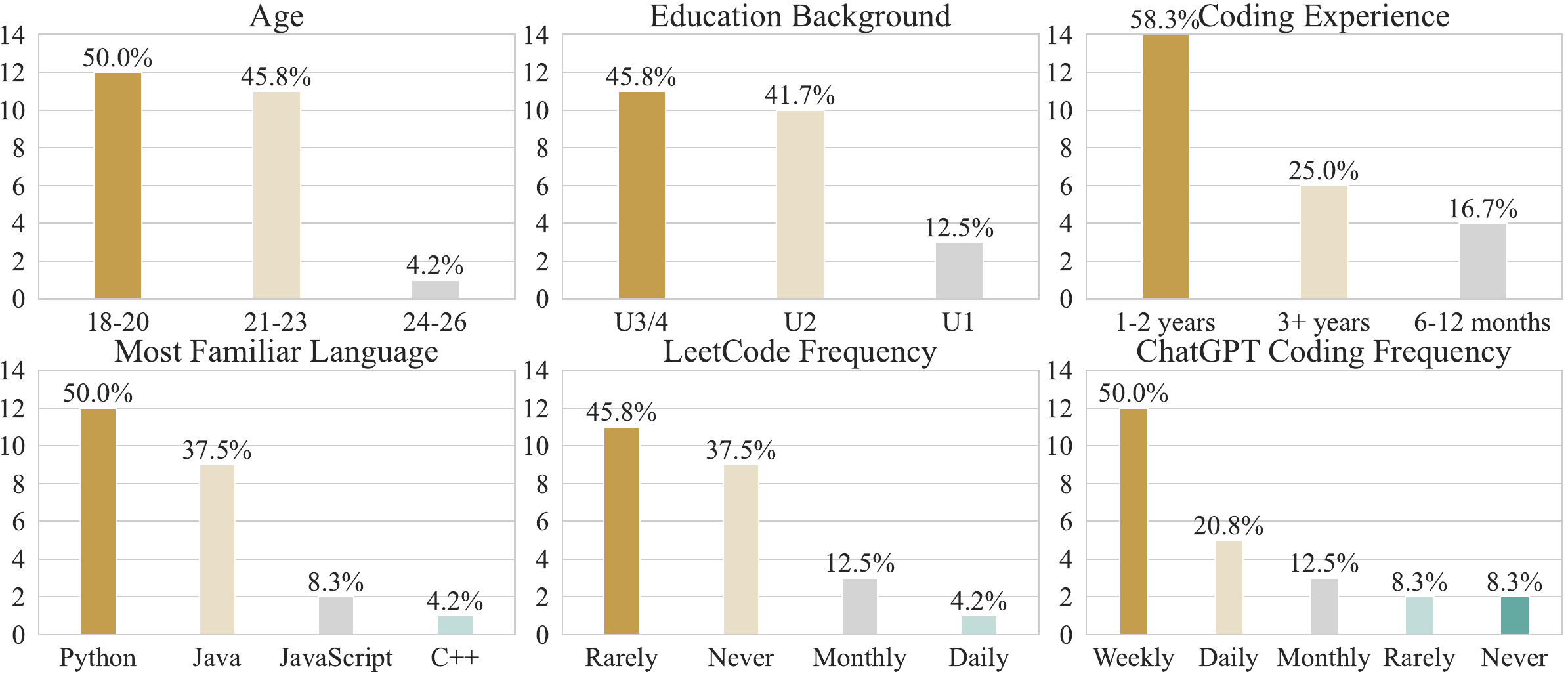}
    \caption{Participant Demographics}
    \label{fig:demo}
\end{figure*}

Figure~\ref{fig:demo} presents the student demographics.
The study involved 24 students, the majority of which are aged 18–23 (23/24, 95.8\%) and enrolled in Electrical, Computer, and Software Engineering or Computer Science programs. \aj{do we have information on which degree programs? SE?CS? Arts? Can we describe/list them here?}\rf{revised}
Most participants reported 1–2 years of coding experience (14/24, 58.3\%) with Python as the most familiar programming language (12/24, 50.0\%), followed by Java (9/24, 37.5\%), with JavaScript and C++ mentioned less frequently.
Participants had completed a range of core computing courses, including Introduction to Programming, Software Systems, Data Structures, Algorithm Design, Operating Systems, and Programming Languages. 
These courses indicate that Beginner students had primarily taken introductory programming and data structure courses, whereas Intermediates had completed more advanced courses such as algorithms, operating systems, and software engineering.\rf{revised}
20 participants (83.3\%) indicated that they rarely or never used LeetCode. 
In contrast, ChatGPT was widely used, with 17 participants (70.8\%) coding with it at least weekly.

\subsection{User Perceptions of ChatGPT (RQ1)}\aj{This is also a section I would expect to see a t-test results with p-values across two different programming experience levels. But if we haven't done them, we can also frame it as a general description of people's perception and not necessarily anything we are trying to hypothesize and empirically support.}

\begin{figure*}[t]
    \centering
    \includegraphics[width=\linewidth]{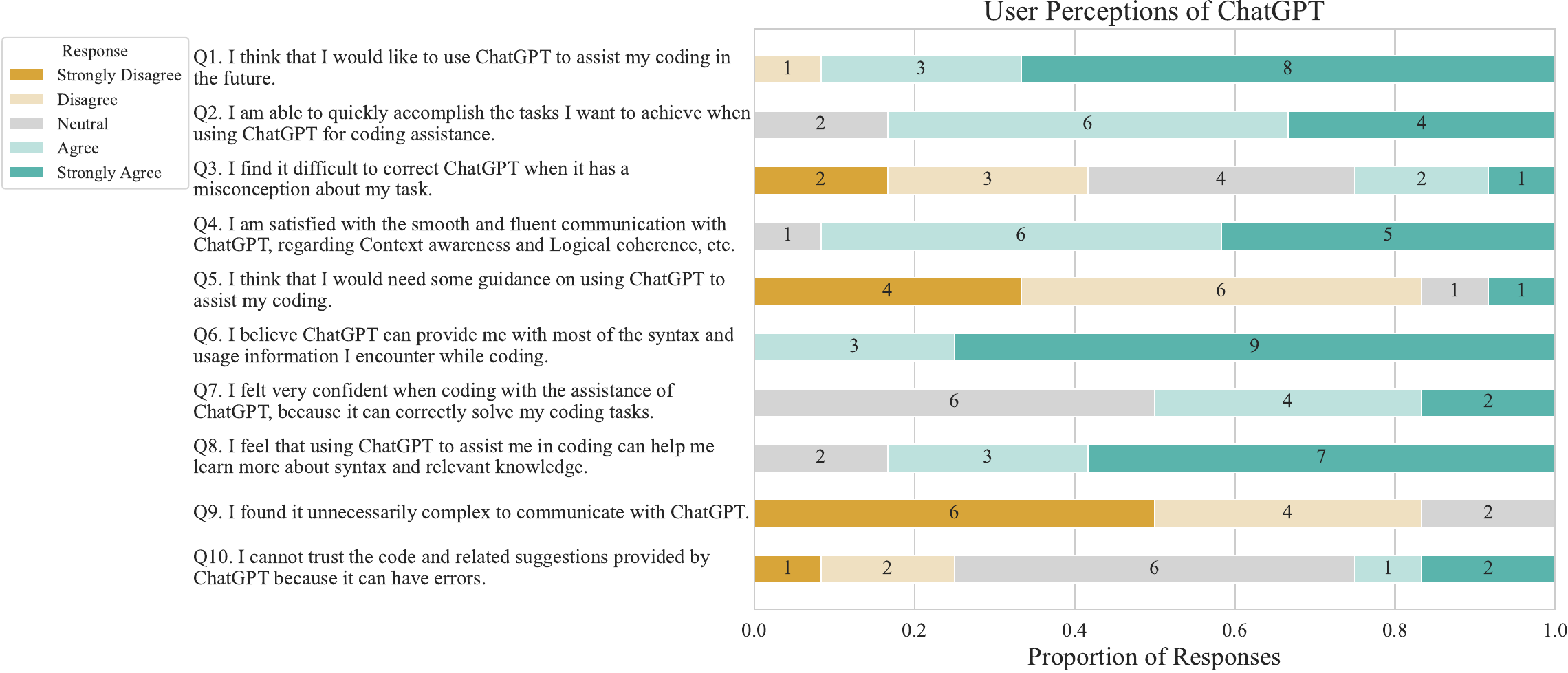}
    \caption{User Perceptions of ChatGPT}
    \label{fig:perceptions}
\end{figure*}

Figure~\ref{fig:perceptions} shows all ten survey questions on user perceptions along with the survey results.

\textbf{Overall Acceptance and Future Use (Q1, Q2)}
Participants generally expressed strong acceptance of ChatGPT. 
Most users either \textit{agreed} or \textit{strongly agreed} that they would like to use ChatGPT in the future (Q1) and that it helps them quickly accomplish their coding goals (Q2). 
These responses indicate positive overall impressions of ChatGPT’s practical value in coding contexts.

\textbf{Communication and Interaction Quality (Q3, Q4, Q9)}
Responses to questions related to interaction quality painted a mixed picture. 
Participants largely \textit{agreed} that ChatGPT communicates smoothly (Q4), suggesting good performance in maintaining context and logical flow. 
Additionally, Q9 responses show that most users did not find ChatGPT unnecessarily complex to communicate with.
However, Q3 revealed that many users found it difficult to correct ChatGPT’s misconceptions, with responses skewed toward \textit{disagree} or \textit{neutral}. 
These three questions together reveal a mixed picture: while communication is mostly smooth, users sometimes struggle with redirecting the model when it goes off-track.

\textbf{Guidance (Q5)}
For the statement on the need for guidance when using ChatGPT (Q5), half of the participants selected \textit{Neutral}, while 4 out of 12 disagreed.
This suggests that most students do not find ChatGPT particularly difficult to use for coding tasks, indicating a relatively low barrier to entry in terms of usability.

\textbf{Syntax Support (Q6, Q8)}
Questions Q6 and Q8 examined whether ChatGPT supports learning and providing syntax-related information. Responses were strongly positive: all participants agreed or strongly agreed in Q6, and 10 out of 12 did so in Q8.
These results suggest that students perceive ChatGPT as a valuable tool not only for coding assistance but also for reinforcing their understanding of programming syntax.

\textbf{Confidence and Trust (Q7, Q10)}
Question Q7 explores students’ confidence when using ChatGPT, while Q10 addresses their trust in its output. 
For Q7, half of the participants expressed confidence in ChatGPT, while the others remained neutral. 
Q10 reflects more hesitation: only three out of 12 participants agreed that they trusted ChatGPT’s code and suggestions, with the majority either disagreeing or remaining neutral. 
This suggests that although students may hold some skepticism about the reliability of ChatGPT’s outputs, they still feel confident using it.
\lili{This is a bit contraversial since the results show that for Q10, most of them are neutral - neither agreeing nor disagreeing}\rf{revised}

To summarize, the participants generally showed a positive attitude towards using ChatGPT for coding tasks, appreciating its usefulness and ease of interaction. 
They reported that ChatGPT helped them accomplish tasks quickly and supported their understanding of syntax. 
While most found communication smooth and intuitive, some struggled with correcting ChatGPT’s misconceptions. 
Notably, while participants expressed concerns about potential errors in ChatGPT’s responses, they generally felt confident using it.
\lili{We may want to revise the last claim.}\rf{revised}

\subsection{Task Performance (RQ2)}


 
Table~\ref{tab:mean_passed_tests} presents the average number of passed tests across four participant groups. 
Interestingly, \textit{Beginner participants with ChatGPT} achieved the highest average score (410.7), substantially surpassing all other groups, including those with higher programming proficiency. 
Their counterparts without ChatGPT scored significantly lower ($p = 0.015$), averaging just 80.7. 
In contrast, the \textit{Intermediate participants}, regardless of ChatGPT access, showed nearly identical performance, with both groups averaging around 291 passed tests. 
This stark contrast suggests that \textit{Beginner} and \textit{Intermediate} participants may have engaged with ChatGPT in fundamentally different ways, leading to divergent outcomes. 
This observation motivates a deeper investigation into how participants interacted with the tool and interpreted its assistance. 

Figure~\ref{fig:tests} shows the distribution of participants in each score range by group.
Participants’ results were grouped into five score ranges: \textit{0}, \textit{1--100}, \textit{100--400}, \textit{400--499}, and \textit{500}. 
An interesting pattern emerges: none of the \textit{Beginner participants without ChatGPT} managed to pass all 500 tests, while every \textit{Intermediate participant with ChatGPT} passed at least one test.
This contrast highlights differences in how groups engaged with the coding tasks and utilized available support.

To further explain these results, we conducted a deeper investigation into participants' conceptual and coding understanding and their detailed chat history with ChatGPT, as discussed in Section~\ref{sec:findings}. \aj{did we do any inferential statistics on this measure? t-test / ANOVA etc? I'd expect to see the results with p-values here.}

\begin{table}[t]
\centering
\caption{Mean Number of Passed Tests per Group}
\begin{tabular}{lr}
\toprule
\textbf{Group} & \textbf{Mean Passed Tests} \\
\midrule
Beginner w/ ChatGPT         & 410.7 \\
Beginner w/o ChatGPT        & 80.7  \\
Intermediate w/ ChatGPT     & 291.0 \\
Intermediate w/o ChatGPT    & 290.8 \\
\midrule
\textbf{All Participants}   & \textbf{268.3} \\
\bottomrule
\end{tabular}
\label{tab:mean_passed_tests}
\end{table}

\begin{figure}[t]
    \centering
    \includegraphics[width=\linewidth]{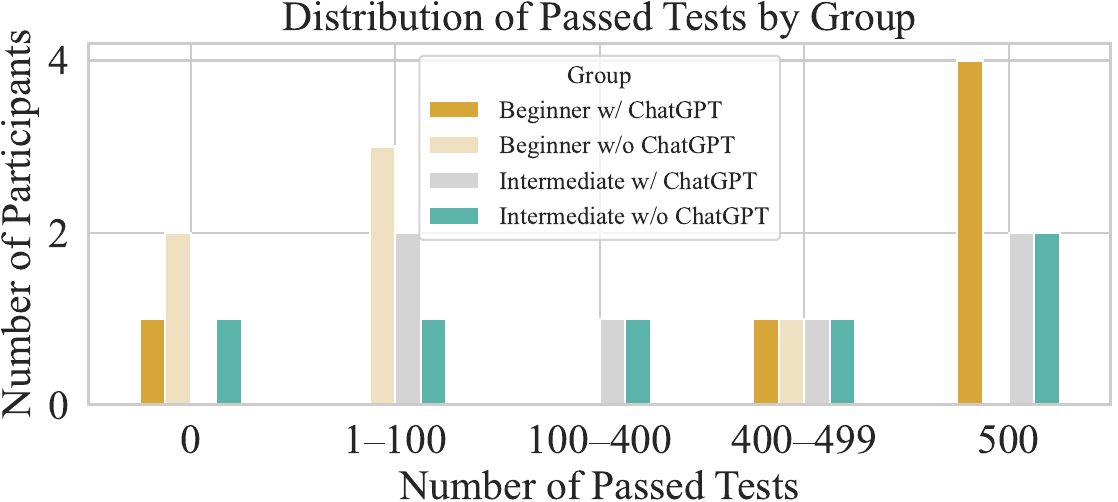}
    \caption{Distribution of Passed Tests by Group \lili{May make the bars shorter}\rf{revised}}
    \label{fig:tests}
\end{figure}

\subsection{Knowledge Gains (RQ3)}

\begin{figure*}[t]
    \centering
    \includegraphics[width=\linewidth]{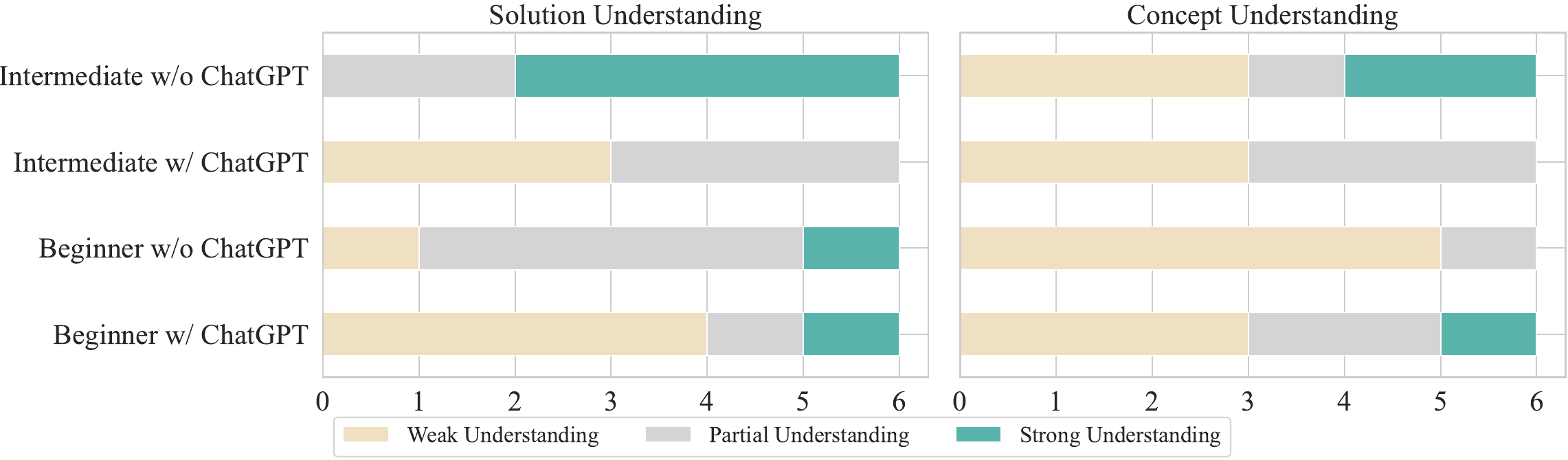}
    \caption{Participants' understanding levels on solution and concept.}
    \label{fig:understand}
\end{figure*}

Figure~\ref{fig:understand} presents the distribution of knowledge gains across four participant groups, with separate plots for solution understanding and conceptual understanding.

For solution understanding, the \textit{Intermediate w/o ChatGPT} group demonstrated the strongest performance, with four out of six participants achieving strong understanding and the remaining two showing partial understanding. 
In contrast, the \textit{Intermediate w/ ChatGPT} group had lower performance, with half showing partial and half showing weak understanding. 
A similar pattern appeared among beginners: while only one \textit{Beginner w/o ChatGPT} participant showed weak understanding, four out of six in the \textit{Beginner w/ ChatGPT} group fell into this category. 
These results indicate that participants without access to ChatGPT generally had a better understanding on their solutions.
This may suggest that when using ChatGPT, some students tend to rely heavily on ChatGPT, which could lead to weaker understanding on their own solutions.

For conceptual understanding, \textit{Intermediate w/o ChatGPT} participants showed stronger overall: three out of six demonstrated strong understanding, and the remaining three showed partial understanding. 
In contrast, among \textit{Intermediate w/ ChatGPT}, half were rated as partial and half as weak, suggesting that \textit{Intermediate w/o ChatGPT} had a better understanding on the concept.
Among beginners, five out of six participants in the \textit{Beginner w/o ChatGPT} group demonstrated partial understanding, and one showed weak understanding. 
Interestingly, \textit{Beginner w/ ChatGPT} participants performed slightly better, with one strong, two partial, and three weak understandings. 
This suggests that, for beginners, ChatGPT may have played a helpful role in understanding the concept.

The contrast in these patterns implies that beginner and intermediate students may interact with ChatGPT in different ways. 
These findings motivate a deeper analysis of how participants used ChatGPT and what differences emerged across proficiency levels.

\subsection{Interaction Patterns with ChatGPT (RQ4)}
\label{sec:interaction}
Table~\ref{tab:chat_history} summarizes interaction patterns from chat histories, including definitions and participants who used each pattern.

Complete Solution Generation was used by three beginners and two intermediates, while the remaining three beginners and four intermediates adopted Stepwise Code Generation, indicating a preference for generating code incrementally rather than generating one-shot solutions.
Code Improvement appeared in three beginners but only one intermediate, suggesting that beginners were more likely to seek assistance in refining or optimizing their code.
Code Explanation was used equally, with two participants each using ChatGPT to understand code behavior. 
API Usage was the most frequent pattern, used by all six beginners and four intermediates to clarify function usage or syntax. 
Concept Explanation was more common among beginners (four participants) than intermediates (one participant), which may help explain the stronger conceptual understanding observed in the beginner group. 
Bug Detection and Fixing was relatively rare, found in just one beginner and one intermediate. Finally, Test Generation was seen in only one beginner.
This shows that students primarily focused on code generation and rarely explored GenAI's broader functionalities, such as debugging or test generation, which could support deeper understanding and better code verification.

These findings reveal behavioral differences between beginner and intermediate participants in how they engaged with ChatGPT.
In particular, beginners often sought conceptual explanations and code refinement support, while intermediates leaned toward stepwise generation.
These patterns may help explain the observed differences in knowledge gains.
To explore this further, we next examine how these interactions relate to participants’ performance and understanding levels.

\begin{table*}[t]
\centering
\caption{Common ChatGPT Interaction Patterns Across Participant Groups}
\label{tab:chat_history}
\renewcommand{\arraystretch}{1.25}
\setlength{\tabcolsep}{2pt} 
\begin{tabular}{@{}p{4cm}p{5.8cm}|*{6}{p{0.5cm}}|*{6}{p{0.5cm}}@{}}
\toprule
\multirow{2}{*}{\textbf{Pattern}} & \multirow{2}{*}{\textbf{Explanation}} & 
\multicolumn{6}{c|}{\textbf{Beginner}} & 
\multicolumn{6}{c}{\textbf{Intermediate}} \\
\cmidrule(lr){3-8} \cmidrule(lr){9-14}
 & & P1 & P2 & P3 & P4 & P5 & P6 & P7 & P8 & P9 & P10 & P11 & P12 \\
\midrule
\rowcolor{gray!10}
Complete Solution Generation    
& Directly copying the entire task prompt into ChatGPT to generate a complete solution
&  \checkmark & \checkmark  & \checkmark  &   &   &  \checkmark &  \checkmark &   &   & \checkmark  &   &   \\
Stepwise Code Generation  
& Inputting code snippets along with comments or subtask descriptions to generate specific parts of the solution                             
&   &  &   & \checkmark  & \checkmark  &  &   & \checkmark  & \checkmark  &    & \checkmark  &  \checkmark \\
\rowcolor{gray!10}
Code Improvement           
& Asking ChatGPT to improve or refactor existing code                            
&   &  \checkmark & \checkmark  &   &   & \checkmark  &   & \checkmark  &   &   &   &   \\
Code Explanation           
& Prompting ChatGPT to explain code behavior or logic                            
& \checkmark  &   &  \checkmark &   &   &   &   & \checkmark  &   &   &   & \checkmark  \\
\rowcolor{gray!10}
API Usage                  
& Queries about API usage or function behavior                                   
& \checkmark  & \checkmark  & \checkmark  & \checkmark  & \checkmark  & \checkmark  &    & \checkmark  &  \checkmark &   & \checkmark  &  \checkmark \\
Concept Explanation        
& Asking for the concept clarification                   
&  \checkmark & \checkmark  &   & \checkmark  &   &   &   &   &   & \checkmark  &   &   \\
\rowcolor{gray!10}
Bug Detection and Fixing   
& Requesting help to debug and correct code errors                               
&   &   & \checkmark  &   &   &   & \checkmark  &   &   &   &   &   \\
Test Generation            
& Asking ChatGPT to write test cases or suggest test strategies                  
&   &   &   &   &   &   & \checkmark  &   &   &   &   &   \\
\bottomrule
\end{tabular}
\end{table*}
\section{Findings}
\label{sec:findings}
In this section, we synthesize the results presented, highlighting key links between participants’ task performance, knowledge gains, and interaction patterns with ChatGPT.

\subsection{Complete Solution Generation vs. Stepwise Code Generation}
As discussed in Section~\ref{sec:interaction}, six participants used Complete Solution Generation, while the other six used Stepwise Code Generation. 
Participants who used Complete Solution Generation achieved an average score of 495, with five out of six passing all tests. 
In comparison, those who used Stepwise Code Generation had a lower average score of 206.7, and only one participant passed all tests.
This difference was statistically significant ($p = 0.032$).
Among these 12 participants, only one (P3), who adopted the Complete Solution Generation strategy, demonstrated a strong understanding.
\lili{We may not mention statitical tests if we do not present more details. Also, it is a bit unclear what exactly are the two groups? Or, if we want to include this in the finding, we may add some details on the statistical test we did. e.g., we conducted xx test to examine xxx.}\rf{revised}

To better understand these results, we examined participants’ code, survey responses, chat transcripts, and screen recordings.
Among those who used Complete Solution Generation, several (P1, P2, P6, and P10) appeared to focus primarily on completing the task rather than understanding and applying the underlying concept. 
For example,although P1 and P2 asked ChatGPT to explain the concept, their inability to understand the generated solution limited their ability to connect it to the concept meaningfully.
In contrast, P3 also used Complete Solution Generation but took a more active learning approach: Ask ChatGPT for detailed explanations and read them carefully. 
As a result, P3 was able to articulate both their solution and the concept effectively.

Among the Stepwise Code Generation group, participants like P4, P8, and P9 attempted to guide ChatGPT through subtasks based on their own approach. 
However, their flawed logic or design lead to errors despite the more interactive interaction process.

\begin{tcolorbox}[left=3pt,right=3pt,top=1pt,bottom=1pt]
	\textbf{Finding 1:}
	\textit{
		Using GenAI to generate complete solutions can significantly improve task performance. 
        However, it does not lead to statistically significant improvements in knowledge gains. 
        For students primarily focused on completing the task, Complete Solution Generation may lead to overreliance on the tool, which can hinder deeper understanding.
	}
\end{tcolorbox}

\subsection{Beginners vs. Intermediates}
\label{sec:gvt}

\textbf{Different impacts of ChatGPT on Task Performance and Knowledge Gains.}
For beginners, as shown in Table~\ref{tab:mean_passed_tests}, those with access to ChatGPT achieved a significantly higher average number of passed tests (410.7) compared to those without (80.7). 
This substantial performance gap can be attributed to two key factors. 
First, four out of six \textit{Beginners w/ ChatGPT} used the Complete Solution Generation strategy, directly obtaining complete code. 
Second, \textit{Beginners w/o ChatGPT}—due to their limited coding experience-often struggled to translate their ideas into functional code. 
For example, the solutions of P14 and P15 had compilation errors.

However, when evaluating solution understanding, the trend reversed. 
Beginners without ChatGPT generally showed stronger understanding. 
Despite difficulties in writing correct code, they explained their reasoning more clearly.
In contrast, many of the \textit{Beginners w/ ChatGPT}, while successful in producing working code, failed to fully grasp the rationale behind it, as evidenced by Figure~\ref{fig:understand} where most of them demonstrated a weak understanding.

For intermediate participants, the task performance was similar across both conditions, indicating that their existing proficiency allowed them to perform similarly with or without ChatGPT. 
However, the knowledge gains revealed differences. 
Among \textit{Intermediate w/o ChatGPT}, four participants got strong understanding of their solutions, and two got strong understanding of the concept. 
On the other hand,  half of \textit{Intermediate w/o ChatGPT} got only partial understanding, and the other half were categorized as weak.

\begin{tcolorbox}[left=3pt,right=3pt,top=1pt,bottom=1pt]
	\textbf{Finding 2:}
	\textit{
        While ChatGPT significantly improved task performance for beginners by helping them overcome implementation challenges, it did not improve their understanding.
        Conversely, intermediates without ChatGPT exhibited stronger conceptual and solution understanding, suggesting that intermediate students may not leverage GenAI effectively to support in-depth learning.
	}
\end{tcolorbox}

\textbf{Differences in Interaction Patterns and Usage Intents.}
Four out of six beginners employed the Complete Solution Generation strategy, compared to only two out of six intermediates, suggesting that beginners tended to rely more heavily on GenAI, while intermediates relied more on their own skills.
This difference in usage strategies may stem from the timing of students' exposure to GenAI.
ChatGPT became widely available in late 2022 ~\cite{ChatGPT}, whereas most intermediate participants in our study began programming in 2021 or earlier.
Beginners may have encountered GenAI before developing solid programming foundations, and therefore lacked experience using conventional resources to solve problems.
As a result, they often viewed GenAI as their primary problem solving tool, which explains their poor performance without it.
In contrast, intermediates likely developed their skills through conventional coursework and had more experience leveraging conventional learning resources.
\lili{This is a bit confusing without introducing the background that the intermediate students entered university before ChatGPT became popular. Do we want to clarify this? Or, do we want to rule out this factor?}\rf{revised}
This background helped them achieve the best performance and knowledge gains in the no-ChatGPT condition.
As a result, beginners were more inclined to delegate full control to the tool, whereas intermediates adopted a more cautious or incremental approach.

\begin{tcolorbox}[left=3pt,right=3pt,top=1pt,bottom=1pt]
	\textbf{Finding 3:}
	\textit{
		Differences in usage intent appear to be influenced by the timing of GenAI exposure relative to programming experience. 
        Beginners, who often encountered GenAI before developing strong programming foundations and had limited experience with conventional online resources, tended to rely on it as a primary aid. 
        In contrast, intermediates, who built their skills through conventional learning methods before using GenAI, integrated it more cautiously into their workflow. 
        As a result, even without ChatGPT, they still achieved substantial knowledge gains.
	}
\end{tcolorbox}

\subsection{Balanced Use of GenAI}

Our study reveals that many students lacked effective strategies for using GenAI, leading to limited gains.
Both over-reliance and minimal use proved ineffective.
These patterns were consistent across beginner and intermediate participants.

For instance, P4 stated, “I didn’t want it to solve the entire problem for me,” yet attempted to implement the solution without fully understanding the concept. 
Similarly, P9 only used ChatGPT for syntax generation, neglecting opportunities to ask questions or explore explanations, resulting in limited knowledge gains.
On the other end of the spectrum, participants like P1, P2, P6, and P10 relied heavily on ChatGPT to complete the task. 
While they produced working code and passed tests, their weak conceptual understanding suggested little true learning had occurred.

We also observed cases like P16 and P24, who showed strong conceptual understanding but lacked the coding skills to pass many test cases. 
This emphasizes that successful learning involves both grasping concepts and acquiring practical implementation skills.

These findings highlight the need for support.
Instructors should not take an all-or-nothing stance, either banning GenAI entirely or allowing unrestricted, unguided use. 
Instead, they should help students use GenAI as a structured learning aid. 

\begin{tcolorbox}[left=3pt,right=3pt,top=1pt,bottom=1pt]
\textbf{Finding 4:}
\textit{
Both over- and under-utilization of GenAI hinder effective learning.
Most participants lacked strategies for using GenAI productively, highlighting the need for clearer guidelines to support its educational use.
}
\end{tcolorbox}

\section{Tips for Students on Using GenAI Effectively}

Our findings suggest that the way students interact with GenAI tools significantly influences both task performance and knowledge gains.
Based on these insights, we present the following tips to help students use GenAI more effectively in programming tasks:

\textbf{Avoid Over-Reliance on Complete Code Generation.}
    Using GenAI to generate full solutions can help students complete a task, but without understanding the logic, learning suffers.
    \textit{
    For example, P1 and P2 both used Complete Solution Generation to successfully pass all tests. 
    However, they failed to develop conceptual understanding, as they could not explain how the code worked when asked.
    }
    Always take time to understand what the AI generated and why.
        
\textbf{Develop a Clear Plan Before Coding.}
    Before initiating code generation, students should first form a coherent plan grounded in their understanding of the task.
    \textit{
    For example, P8 and P9 used Stepwise Code Generation by asking ChatGPT to help with individual parts of their solutions. 
    However, their initial approaches were based on flawed assumptions, which led to poor task performance.
    }
    Starting with personal reasoning is valuable, but if uncertain, it’s recommended to consult GenAI for conceptual explanations—not just code—to validate the approach and avoid carrying early misunderstandings forward.

\textbf{Use GenAI as a Learning Partner, Not Just a Tool.} 
    GenAI is most effective when used to complement students' own thinking—not to fully replace it or be used too minimally.
    \textit{
    Participants like P2, P6, and P7 primarily used GenAI to generate code and focused only on completing the task. 
    As a result, they showed limited understanding on both their code and the concept. 
    On the other hand, P4 tried to solve the task independently without asking for clarification, despite not fully understanding the concept. 
    This led to poor performance and weak understanding.
    }
    Notably, only 33\% of participants asked ChatGPT for code or conceptual explanations.
    To benefit more fully, students should share their reasoning with ChatGPT, ask for clarification when uncertain, and actively engage with its explanations—not just its output.

\textbf{Explore Beyond Code Generation.}
    GenAI can support more than code generation—it also aids bug detection and test generation, though few participants (only P3 and P7) used it this way.
    Using GenAI to identify bugs can uncover flawed assumptions and prompt students to revisit the underlying concept.
    Similarly, requesting test generation helps ensure solutions handle edge cases—not just the obvious ones.
    These features help uncover assumptions, verify logic, and strengthen understanding, making them valuable for long-term learning.

By following these tips, students can maximize the benefits of GenAI in programming education, leveraging its strengths for both task completion and deeper learning.
\section{Conclusion}
We conducted a controlled study with 24 undergraduate students to examine how learners with different proficiency levels use GenAI to solve programming tasks.
Our goal was to understand GenAI’s impact on both task performance and conceptual understanding, and to identify strategies that support meaningful learning.

Through analysis of knowledge gains and interactions, we found that while using GenAI to generate complete solutions can significantly improve task performance, it does not necessarily enhance knowledge gains. 
Beginners benefited from ChatGPT in completing the task, yet often lacked deeper understanding. 
Intermediates, on the other hand, performed better without ChatGPT, suggesting they may not yet know how to use the tool effectively for learning. 
We also observed that when and how students were first introduced to GenAI shaped their usage intent. 
Finally, both over-reliance and minimal use of GenAI were found to hinder learning, pointing to a widespread need for better usage strategies.

Based on these findings, we outlined tips for students and educators to support more effective use of GenAI as a learning companion.
Our work contributes to the growing understanding of how GenAI fits into programming education and underscores the importance of guidance in its integration.

\section*{Acknowledgements}
This work is supported by the Natural Sciences and Engineering Research Council of Canada Discovery Grant (Grant No. RGCPIN-2022-03744 and Grant No. DGECR-2022-00378), 
McGill eLATE Teaching and Learning Improvement Funds 2023-2024,
and Fonds de recherche du Québec-secteur Nature et technologies (Grant No.363482~\cite{frqntNewAca}).
We thank McGill University for supporting this study through its 
Research Ethics Board (REB) review and compliance procedures.
We are deeply grateful to all participants who took part in our user study 
and generously shared their time and insights, which made this work possible.

\balance
\bibliographystyle{IEEEtran}
\bibliography{References/Papers,References/Links}

@misc{randoop,
    url     =   {{https://randoop.github.io/randoop/}},
    title   =   {Randoop - Automatic unit test generation for Java},
    year    =   {2025}
}

@misc{ChatGPT,
    url     =   {{https://openai.com/index/chatgpt/}},
    title   =   {OpenAI - Introducing ChatGPT},
    year    =   {2022}
}

@misc{replication_package,
    url     =   {{https://github.com/RafferyChen/Examining-the-Usage-of-Generative-AI-Models-in-Student-Learning.git}},
    title   =   {Examining-the-Usage-of-Generative-AI-Models-in-Student-Learning},
    year    =   {2025}
}

@inproceedings{10.1145/3511861.3511863,
author = {Finnie-Ansley, James and Denny, Paul and Becker, Brett A. and Luxton-Reilly, Andrew and Prather, James},
title = {The Robots Are Coming: Exploring the Implications of OpenAI Codex on Introductory Programming},
year = {2022},
isbn = {9781450396431},
publisher = {Association for Computing Machinery},
address = {New York, NY, USA},
url = {https://doi.org/10.1145/3511861.3511863},
doi = {10.1145/3511861.3511863},
booktitle = {Proceedings of the 24th Australasian Computing Education Conference},
pages = {10–19},
numpages = {10},
location = {Virtual Event, Australia},
series = {ACE '22}
}

@inproceedings{10.1145/3576123.3576134,
author = {Finnie-Ansley, James and Denny, Paul and Luxton-Reilly, Andrew and Santos, Eddie Antonio and Prather, James and Becker, Brett A.},
title = {My AI Wants to Know if This Will Be on the Exam: Testing OpenAI’s Codex on CS2 Programming Exercises},
year = {2023},
isbn = {9781450399418},
publisher = {Association for Computing Machinery},
address = {New York, NY, USA},
url = {https://doi.org/10.1145/3576123.3576134},
doi = {10.1145/3576123.3576134},
booktitle = {Proceedings of the 25th Australasian Computing Education Conference},
pages = {97–104},
numpages = {8},
location = {Melbourne, VIC, Australia},
series = {ACE '23}
}

@inproceedings{10.1145/3587102.3588814,
author = {Cipriano, Bruno Pereira and Alves, Pedro},
title = {GPT-3 vs Object Oriented Programming Assignments: An Experience Report},
year = {2023},
isbn = {9798400701382},
publisher = {Association for Computing Machinery},
address = {New York, NY, USA},
url = {https://doi.org/10.1145/3587102.3588814},
doi = {10.1145/3587102.3588814},
booktitle = {Proceedings of the 2023 Conference on Innovation and Technology in Computer Science Education V. 1},
pages = {61–67},
numpages = {7},
location = {Turku, Finland},
series = {ITiCSE 2023}
}

@inproceedings{10.1145/3545945.3569770,
author = {Leinonen, Juho and Hellas, Arto and Sarsa, Sami and Reeves, Brent and Denny, Paul and Prather, James and Becker, Brett A.},
title = {Using Large Language Models to Enhance Programming Error Messages},
year = {2023},
isbn = {9781450394314},
publisher = {Association for Computing Machinery},
address = {New York, NY, USA},
url = {https://doi.org/10.1145/3545945.3569770},
doi = {10.1145/3545945.3569770},
booktitle = {Proceedings of the 54th ACM Technical Symposium on Computer Science Education V. 1},
pages = {563–569},
numpages = {7},
location = {Toronto ON, Canada},
series = {SIGCSE 2023}
}

@inproceedings{10.1145/3587102.3588805,
author = {Reeves, Brent and Sarsa, Sami and Prather, James and Denny, Paul and Becker, Brett A. and Hellas, Arto and Kimmel, Bailey and Powell, Garrett and Leinonen, Juho},
title = {Evaluating the Performance of Code Generation Models for Solving Parsons Problems With Small Prompt Variations},
year = {2023},
isbn = {9798400701382},
publisher = {Association for Computing Machinery},
address = {New York, NY, USA},
url = {https://doi.org/10.1145/3587102.3588805},
doi = {10.1145/3587102.3588805},
booktitle = {Proceedings of the 2023 Conference on Innovation and Technology in Computer Science Education V. 1},
pages = {299–305},
numpages = {7},
location = {Turku, Finland},
series = {ITiCSE 2023}
}

@misc{dakhel2023githubcopilotaipair,
      title={GitHub Copilot AI pair programmer: Asset or Liability?}, 
      author={Arghavan Moradi Dakhel and Vahid Majdinasab and Amin Nikanjam and Foutse Khomh and Michel C. Desmarais and Zhen Ming and Jiang},
      year={2023},
      eprint={2206.15331},
      archivePrefix={arXiv},
      primaryClass={cs.SE},
      url={https://arxiv.org/abs/2206.15331}, 
}

@misc{denny2023promptlyusingpromptproblems,
      title={Promptly: Using Prompt Problems to Teach Learners How to Effectively Utilize AI Code Generators}, 
      author={Paul Denny and Juho Leinonen and James Prather and Andrew Luxton-Reilly and Thezyrie Amarouche and Brett A. Becker and Brent N. Reeves},
      year={2023},
      eprint={2307.16364},
      archivePrefix={arXiv},
      primaryClass={cs.HC},
      url={https://arxiv.org/abs/2307.16364}, 
}

@article{10.1145/3715762,
author = {Rahe, Christian and Maalej, Walid},
title = {How Do Programming Students Use Generative AI?},
year = {2025},
issue_date = {July 2025},
publisher = {Association for Computing Machinery},
address = {New York, NY, USA},
volume = {2},
number = {FSE},
url = {https://doi.org/10.1145/3715762},
doi = {10.1145/3715762},
journal = {Proc. ACM Softw. Eng.},
month = jun,
articleno = {FSE045},
numpages = {23}
}

@article{10.1145/3617367,
author = {Prather, James and Reeves, Brent N. and Denny, Paul and Becker, Brett A. and Leinonen, Juho and Luxton-Reilly, Andrew and Powell, Garrett and Finnie-Ansley, James and Santos, Eddie Antonio},
title = {“It’s Weird That it Knows What I Want”: Usability and Interactions with Copilot for Novice Programmers},
year = {2023},
issue_date = {February 2024},
publisher = {Association for Computing Machinery},
address = {New York, NY, USA},
volume = {31},
number = {1},
issn = {1073-0516},
url = {https://doi-org.proxy3.library.mcgill.ca/10.1145/3617367},
doi = {10.1145/3617367},
journal = {ACM Trans. Comput.-Hum. Interact.},
month = nov,
articleno = {4},
numpages = {31}
}

@inproceedings{10.1145/3544548.3580919,
author = {Kazemitabaar, Majeed and Chow, Justin and Ma, Carl Ka To and Ericson, Barbara J. and Weintrop, David and Grossman, Tovi},
title = {Studying the effect of AI Code Generators on Supporting Novice Learners in Introductory Programming},
year = {2023},
isbn = {9781450394215},
publisher = {Association for Computing Machinery},
address = {New York, NY, USA},
url = {https://doi-org.proxy3.library.mcgill.ca/10.1145/3544548.3580919},
doi = {10.1145/3544548.3580919},
booktitle = {Proceedings of the 2023 CHI Conference on Human Factors in Computing Systems},
articleno = {455},
numpages = {23},
location = {Hamburg, Germany},
series = {CHI '23}
}

@article{Sivasakthi2025,
  author    = {M. Sivasakthi and A. Meenakshi},
  title     = {Generative AI in Programming Education: Evaluating ChatGPT’s Effect on Computational Thinking},
  journal   = {SN Computer Science},
  year      = {2025},
  volume    = {6},
  number    = {5},
  pages     = {541},
  doi       = {10.1007/s42979-025-04051-9},
  url       = {https://doi.org/10.1007/s42979-025-04051-9},
  issn      = {2661-8907}
}

@inproceedings{10.1145/3719160.3736625,
author = {Penney, Jacob and Acharya, Pawan and Hilbert, Peter and Parekh, Priyanka and Sarma, Anita and Steinmacher, Igor and Gerosa, Marco Aurelio},
title = {Outcomes, Perceptions, and Interaction Strategies of Novice Programmers Studying with ChatGPT},
year = {2025},
isbn = {9798400715273},
publisher = {Association for Computing Machinery},
address = {New York, NY, USA},
url = {https://doi.org/10.1145/3719160.3736625},
doi = {10.1145/3719160.3736625},
booktitle = {Proceedings of the 7th ACM Conference on Conversational User Interfaces},
articleno = {73},
numpages = {15},
location = {
},
series = {CUI '25}
}

@inproceedings{10.1145/3613904.3642706,
author = {Nguyen, Sydney and Babe, Hannah McLean and Zi, Yangtian and Guha, Arjun and Anderson, Carolyn Jane and Feldman, Molly Q},
title = {How Beginning Programmers and Code LLMs (Mis)read Each Other},
year = {2024},
isbn = {9798400703300},
publisher = {Association for Computing Machinery},
address = {New York, NY, USA},
url = {https://doi-org.proxy3.library.mcgill.ca/10.1145/3613904.3642706},
doi = {10.1145/3613904.3642706},
booktitle = {Proceedings of the 2024 CHI Conference on Human Factors in Computing Systems},
articleno = {651},
numpages = {26},
location = {Honolulu, HI, USA},
series = {CHI '24}
}

@article{2021Chen,
author = {Chen, Mark and Tworek, Jerry and Jun, Heewoo and Yuan, Qiming and Ponde, Henrique and Kaplan, Jared and Edwards, Harri and Burda, Yura and Joseph, Nicholas and Brockman, Greg and Ray, Alex and Puri, Raul and Krueger, Gretchen and Petrov, Michael and Khlaaf, Heidy and Sastry, Girish and Mishkin, Pamela and Chan, Brooke and Gray, Scott and Zaremba, Wojciech},
year = {2021},
month = {07},
pages = {},
title = {Evaluating Large Language Models Trained on Code},
doi = {10.48550/arXiv.2107.03374}
}

@article{Singhal2023,
  author = {Singhal, Karan and Azizi, Shekoofeh and Tu, Tao and Mahdavi, S. Sara and Wei, Jason and Chung, Hyung Won and Scales, Nathan and Tanwani, Ajay and Cole-Lewis, Heather and Pfohl, Stephen and Payne, Perry and Seneviratne, Martin and Gamble, Paul and Kelly, Chris and Babiker, Abubakr and Schärli, Nathanael and Chowdhery, Aakanksha and Mansfield, Philip and Demner-Fushman, Dina and Agüera y Arcas, Blaise and Webster, Dale and Corrado, Greg S. and Matias, Yossi and Chou, Katherine and Gottweis, Juraj and Tomasev, Nenad and Liu, Yun and Rajkomar, Alvin and Barral, Joelle and Semturs, Christopher and Karthikesalingam, Alan and Natarajan, Vivek},
  title = {Large language models encode clinical knowledge},
  journal = {Nature},
  volume = {620},
  number = {7972},
  pages = {172--180},
  year = {2023},
  month = aug,
  doi = {10.1038/s41586-023-06291-2},
  url = {https://doi.org/10.1038/s41586-023-06291-2},
  issn = {1476-4687}
}

@inproceedings{10.1145/3491101.3519665,
author = {Vaithilingam, Priyan and Zhang, Tianyi and Glassman, Elena L.},
title = {Expectation vs.\&nbsp;Experience: Evaluating the Usability of Code Generation Tools Powered by Large Language Models},
year = {2022},
isbn = {9781450391566},
publisher = {Association for Computing Machinery},
address = {New York, NY, USA},
url = {https://doi.org/10.1145/3491101.3519665},
doi = {10.1145/3491101.3519665},
booktitle = {Extended Abstracts of the 2022 CHI Conference on Human Factors in Computing Systems},
articleno = {332},
numpages = {7},
location = {New Orleans, LA, USA},
series = {CHI EA '22}
}

@article{KASNECI2023102274,
title = {ChatGPT for good? On opportunities and challenges of large language models for education},
journal = {Learning and Individual Differences},
volume = {103},
pages = {102274},
year = {2023},
issn = {1041-6080},
doi = {https://doi.org/10.1016/j.lindif.2023.102274},
url = {https://www.sciencedirect.com/science/article/pii/S1041608023000195},
author = {Enkelejda Kasneci and Kathrin Sessler and Stefan Küchemann and Maria Bannert and Daryna Dementieva and Frank Fischer and Urs Gasser and Georg Groh and Stephan Günnemann and Eyke Hüllermeier and Stephan Krusche and Gitta Kutyniok and Tilman Michaeli and Claudia Nerdel and Jürgen Pfeffer and Oleksandra Poquet and Michael Sailer and Albrecht Schmidt and Tina Seidel and Matthias Stadler and Jochen Weller and Jochen Kuhn and Gjergji Kasneci}
}

@inproceedings{10.1145/3544548.3581225,
author = {Mirowski, Piotr and Mathewson, Kory W. and Pittman, Jaylen and Evans, Richard},
title = {Co-Writing Screenplays and Theatre Scripts with Language Models: Evaluation by Industry Professionals},
year = {2023},
isbn = {9781450394215},
publisher = {Association for Computing Machinery},
address = {New York, NY, USA},
url = {https://doi-org.proxy3.library.mcgill.ca/10.1145/3544548.3581225},
doi = {10.1145/3544548.3581225},
booktitle = {Proceedings of the 2023 CHI Conference on Human Factors in Computing Systems},
articleno = {355},
numpages = {34},
location = {Hamburg, Germany},
series = {CHI '23}
}

@misc{reif2022recipearbitrarytextstyle,
      title={A Recipe For Arbitrary Text Style Transfer with Large Language Models}, 
      author={Emily Reif and Daphne Ippolito and Ann Yuan and Andy Coenen and Chris Callison-Burch and Jason Wei},
      year={2022},
      eprint={2109.03910},
      archivePrefix={arXiv},
      primaryClass={cs.CL},
      url={https://arxiv.org/abs/2109.03910}, 
}

@misc{liu2023generativediscotexttovideogeneration,
      title={Generative Disco: Text-to-Video Generation for Music Visualization}, 
      author={Vivian Liu and Tao Long and Nathan Raw and Lydia Chilton},
      year={2023},
      eprint={2304.08551},
      archivePrefix={arXiv},
      primaryClass={cs.HC},
      url={https://arxiv.org/abs/2304.08551}, 
}

@inproceedings{10.1145/3593663.3593695,
author = {Dobslaw, Felix and Bergh, Peter},
title = {Experiences with Remote Examination Formats in Light of GPT-4},
year = {2023},
isbn = {9781450399562},
publisher = {Association for Computing Machinery},
address = {New York, NY, USA},
url = {https://doi.org/10.1145/3593663.3593695},
doi = {10.1145/3593663.3593695},
booktitle = {Proceedings of the 5th European Conference on Software Engineering Education},
pages = {220–225},
numpages = {6},
location = {Seeon/Bavaria, Germany},
series = {ECSEE '23}
}

@inproceedings{10.1145/3615335.3623035,
author = {York, Eric},
title = {Evaluating ChatGPT: Generative AI in UX Design and Web Development Pedagogy},
year = {2023},
isbn = {9798400703362},
publisher = {Association for Computing Machinery},
address = {New York, NY, USA},
url = {https://doi.org/10.1145/3615335.3623035},
doi = {10.1145/3615335.3623035},
booktitle = {Proceedings of the 41st ACM International Conference on Design of Communication},
pages = {197–201},
numpages = {5},
location = {Orlando, FL, USA},
series = {SIGDOC '23}
}

@inproceedings{10.1145/3636243.3636263,
author = {Feng, Tony Haoran and Denny, Paul and Wuensche, Burkhard and Luxton-Reilly, Andrew and Hooper, Steffan},
title = {More Than Meets the AI: Evaluating the performance of GPT-4 on Computer Graphics assessment questions},
year = {2024},
isbn = {9798400716195},
publisher = {Association for Computing Machinery},
address = {New York, NY, USA},
url = {https://doi.org/10.1145/3636243.3636263},
doi = {10.1145/3636243.3636263},
booktitle = {Proceedings of the 26th Australasian Computing Education Conference},
pages = {182–191},
numpages = {10},
location = {Sydney, NSW, Australia},
series = {ACE '24}
}

@inproceedings{10.1145/3626252.3630938,
author = {Liu, Rongxin and Zenke, Carter and Liu, Charlie and Holmes, Andrew and Thornton, Patrick and Malan, David J.},
title = {Teaching CS50 with AI: Leveraging Generative Artificial Intelligence in Computer Science Education},
year = {2024},
isbn = {9798400704239},
publisher = {Association for Computing Machinery},
address = {New York, NY, USA},
url = {https://doi.org/10.1145/3626252.3630938},
doi = {10.1145/3626252.3630938},
booktitle = {Proceedings of the 55th ACM Technical Symposium on Computer Science Education V. 1},
pages = {750–756},
numpages = {7},
location = {Portland, OR, USA},
series = {SIGCSE 2024}
}

@inproceedings{10.1145/3636243.3636252,
author = {Jury, Breanna and Lorusso, Angela and Leinonen, Juho and Denny, Paul and Luxton-Reilly, Andrew},
title = {Evaluating LLM-generated Worked Examples in an Introductory Programming Course},
year = {2024},
isbn = {9798400716195},
publisher = {Association for Computing Machinery},
address = {New York, NY, USA},
url = {https://doi.org/10.1145/3636243.3636252},
doi = {10.1145/3636243.3636252},
booktitle = {Proceedings of the 26th Australasian Computing Education Conference},
pages = {77–86},
numpages = {10},
location = {Sydney, NSW, Australia},
series = {ACE '24}
}

@article{LEE2024100253,
title = {Cheating in the age of generative AI: A high school survey study of cheating behaviors before and after the release of ChatGPT},
journal = {Computers and Education: Artificial Intelligence},
volume = {7},
pages = {100253},
year = {2024},
issn = {2666-920X},
doi = {https://doi.org/10.1016/j.caeai.2024.100253},
url = {https://www.sciencedirect.com/science/article/pii/S2666920X24000560},
author = {Victor R. Lee and Denise Pope and Sarah Miles and Rosalía C. Zárate}
}

@inproceedings{bhalerao,
author = {Bhalerao, Rasika},
year = {2024},
month = {03},
pages = {1574-1575},
title = {My Learnings from Allowing Large Language Models in Introductory Computer Science Classes},
doi = {10.1145/3626253.3635511}
}

@inproceedings{10.1145/3638067.3638100,
author = {Freire, Andr\'{e} Pimenta and Cardoso, Paula Christina Figueira and Salgado, Andr\'{e} de Lima},
title = {May We Consult ChatGPT in Our Human-Computer Interaction Written Exam? An Experience Report After a Professor Answered Yes},
year = {2024},
isbn = {9798400717154},
publisher = {Association for Computing Machinery},
address = {New York, NY, USA},
url = {https://doi.org/10.1145/3638067.3638100},
doi = {10.1145/3638067.3638100},
booktitle = {Proceedings of the XXII Brazilian Symposium on Human Factors in Computing Systems},
articleno = {6},
numpages = {11},
location = {Macei\'{o}, Brazil},
series = {IHC '23}
}

@inproceedings{10.1145/3622780.3623648,
author = {Kuramitsu, Kimio and Obara, Yui and Sato, Miyu and Obara, Momoka},
title = {KOGI: A Seamless Integration of ChatGPT into Jupyter Environments for Programming Education},
year = {2023},
isbn = {9798400703904},
publisher = {Association for Computing Machinery},
address = {New York, NY, USA},
url = {https://doi.org/10.1145/3622780.3623648},
doi = {10.1145/3622780.3623648},
booktitle = {Proceedings of the 2023 ACM SIGPLAN International Symposium on SPLASH-E},
pages = {50–59},
numpages = {10},
location = {Cascais, Portugal},
series = {SPLASH-E 2023}
}

@inproceedings{10.1145/3643795.3648379,
author = {Rasnayaka, Sanka and Wang, Guanlin and Shariffdeen, Ridwan and Iyer, Ganesh Neelakanta},
title = {An Empirical Study on Usage and Perceptions of LLMs in a Software Engineering Project},
year = {2024},
isbn = {9798400705793},
publisher = {Association for Computing Machinery},
address = {New York, NY, USA},
url = {https://doi.org/10.1145/3643795.3648379},
doi = {10.1145/3643795.3648379},
booktitle = {Proceedings of the 1st International Workshop on Large Language Models for Code},
pages = {111–118},
numpages = {8},
location = {Lisbon, Portugal},
series = {LLM4Code '24}
}

@INPROCEEDINGS{11028221,
  author={Naman, Agrawal and Shariffdeen, Ridwan and Wang, Guanlin and Rasnayaka, Sanka and Iyer, Ganesh Neelakanta},
  booktitle={2025 IEEE/ACM International Workshop on Large Language Models for Code (LLM4Code)}, 
  title={Analysis of Student-LLM Interaction in a Software Engineering Project}, 
  year={2025},
  volume={},
  number={},
  pages={112-119},
  doi={10.1109/LLM4Code66737.2025.00019}
}

@INPROCEEDINGS{10430054,
  author={Liu, Jinrun and Tang, Xinyu and Li, Linlin and Chen, Panpan and Liu, Yepang},
  booktitle={2023 IEEE 23rd International Conference on Software Quality, Reliability, and Security Companion (QRS-C)}, 
  title={ChatGPT vs. Stack Overflow: An Exploratory Comparison of Programming Assistance Tools}, 
  year={2023},
  volume={},
  number={},
  pages={364-373},
  doi={10.1109/QRS-C60940.2023.00105}}

@misc{frqntNewAca,
title={Combiner l'analyse statique et la génération de tests pour une assurance qualité améliorée : Une exploration des problèmes de compatibilité Android
},
year={2025},
doi={10.69777/363482},
}

\end{document}